# Potential for allocative harm in an environmental justice data tool

Benjamin Q. Huynh[1]*†, Elizabeth T. Chin[23]†, Allison Koenecke[4], Derek Ouyang[5], Daniel E. Ho[567], Mathew V. Kiang[1]‡, David H. Rehkopf[1]‡

[1] Department of Epidemiology & Population Health, Stanford University School of Medicine; Stanford, CA, USA
[2] Department of Biomedical Data Science, Stanford University School of Medicine; Stanford, CA, USA
[3] Department of Health Policy, Stanford University School of Medicine; Stanford, CA, USA
[4] Department of Information Science, Cornell University; Ithaca, NY, USA
[5] RegLab, Stanford University; Stanford, CA, USA
[6] Stanford Law School; Stanford, CA, USA
[7] Department of Political Science, Stanford University; Stanford, CA, USA
* Corresponding author. Email: benhuynh@stanford.edu
†These authors contributed equally to this work.
‡These authors contributed equally to this work.

**Abstract:** Neighborhood-level screening algorithms are increasingly being deployed to inform policy decisions. We evaluate one such algorithm, CalEnviroScreen – designed to promote environmental justice and used to guide hundreds of millions of dollars in public funding annually – assessing its potential for allocative harm. We observe the model to be sensitive to subjective model decisions, with 16% of tracts potentially changing designation, as well as financially consequential, estimating the effect of its positive designations as a 104% (62-145%) increase in funding, equivalent to $2.08 billion ($1.56-2.41 billion) over four years. We also observe allocative tradeoffs and susceptibility to manipulation, raising ethical concerns. We recommend incorporating sensitivity analyses to mitigate allocative harm and accountability mechanisms to prevent misuse.

**Main Text:**

Algorithms based on high-resolution data are being increasingly used for a wide range of high-impact use-cases, including policymaking. Many such algorithms have fallen under scrutiny, as audits of them have yielded evidence of allocative harm by disproportionately affecting marginalized populations (*1*, *2*). In particular, area-based measures to identify disadvantaged neighborhoods have recently become widespread for tasks such as allocating vaccines, assessing social vulnerability, and healthcare cost adjustment, but their potential for allocative harm is not well understood (*3–6*).

The California Community Environmental Health Screening Tool (CalEnviroScreen) is a data tool that designates neighborhoods as eligible for capital projects and social services funding, intended to promote environmental justice. CalEnviroScreen's model output is used to designate "disadvantaged communities", for which 25% of proceeds from the state's cap-and-trade program are earmarked (equivalent to an estimated $525 million of earmarked funds in 2021). CalEnviroScreen also directly influences funding from a variety of public and private sources, and is reported to have directed an estimated $12.7 billion in funding (*7*). The funding targets of the tool are varied, including programs for affordable housing, land use strategies, agricultural subsidies, wildfire risk reduction, public transit, and renewable energy. Similar data tools are in use or development at the federal and state levels across the United States of America.

CalEnviroScreen ranks each census tract in the state according to its level of marginalization from the perspective of environmental justice. The algorithm does so by aggregating publicly available tract-level data into a single score, based on variables from four categories: environmental exposures, environmental effects, sensitive populations, and socioeconomic factors. Tracts in the top 25% of scores are designated as disadvantaged communities, representing approximately 10 million residents for whom earmarked funding is made available.

Audits of large-scale algorithms are often hindered by proprietary datasets and opaque methodology, limited to observing black box outputs (*1*, *2*, *8*). By contrast, CalEnviroScreen's transparency is highly unique among similarly impactful algorithms, lending itself to thorough auditing. Here we investigate CalEnviroScreen's inner workings, characterizing model sensitivity, funding impact, ethical concerns, and potential harm-reducing strategies.

**Model sensitivity and funding impact**

The CalEnviroScreen model is highly sensitive to change: we found 16.1% of all tracts could change designation based on small alterations to the model; this represents high designation variation given that only 25% of all tracts receive designation (Fig. 1, Table S1). These large fluctuations in designation are solely due to varying subjective model specifications such as health metrics, pre-processing, and aggregation methods (*9*). For example, changing pre-processing methods – switching from percentile ranking to a more commonly used method like *z*-score standardization – led to a 5.3% change in designated tracts.

In the absence of a ground truth variable, model sensitivity can be a proxy for uncertainty, and assessing it helps identify strengths and weaknesses of the model (*9*). For example, we observe high levels of model sensitivity at the designation threshold (75th percentile), where the predicted tract ranking could vary across models by 44 percentile ranks (Fig. 1). Even tracts as lowly-ranked as the bottom 5th percentile could be eligible under slightly different models. Conversely, we observe lower model sensitivity at the 95th percentile, where the predicted range is 18 percentile ranks. Given this variability in ranking certainty, dichotomizing designation may present a false sense of precision, leading to funding decisions based on unstable information.

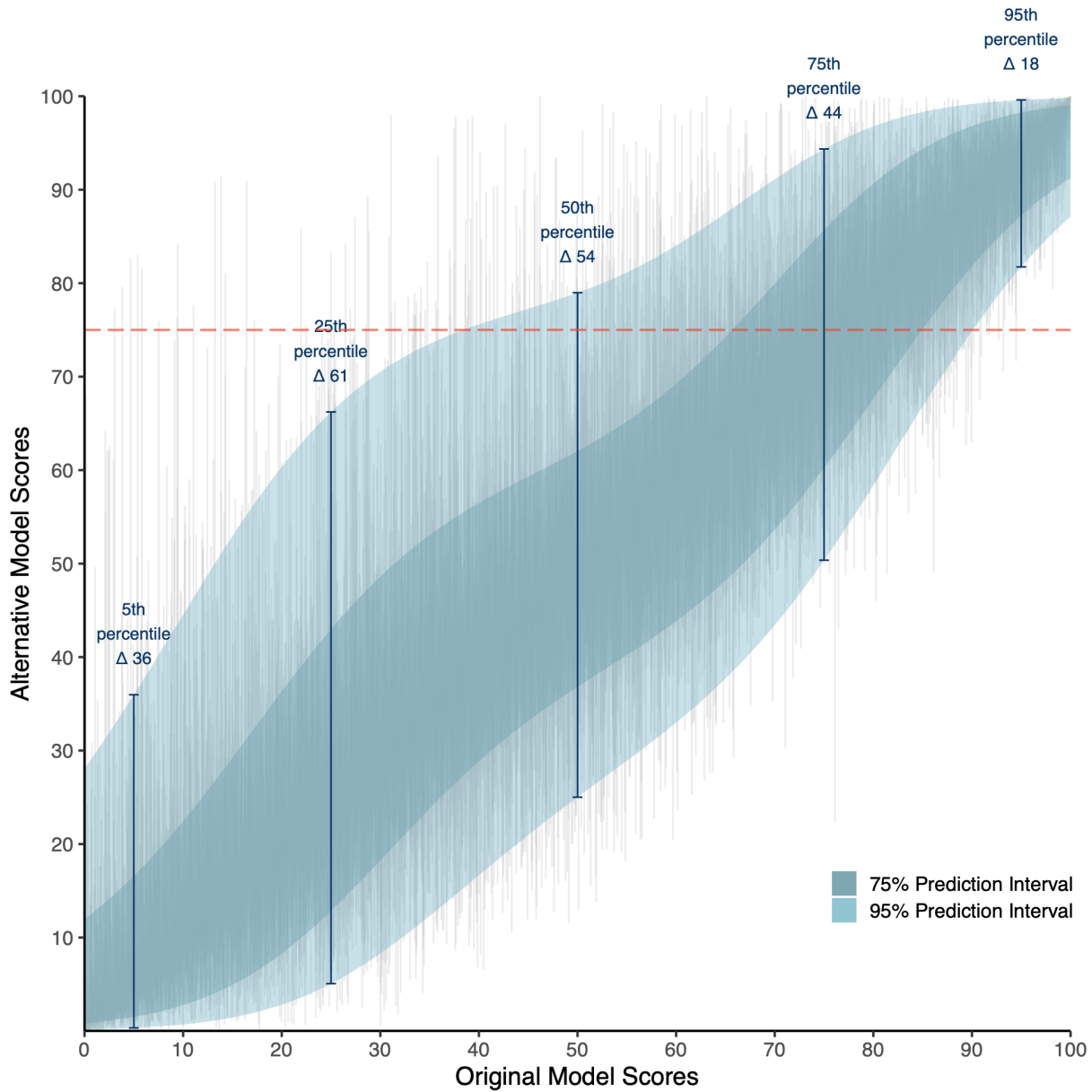

**Fig. 1. CalEnviroScreen's sensitivity to input parameters**. Axes denote model scores in terms of percentiles. Gray bars indicate maximum and minimum values from alternative plausible model specifications varying health metrics, pre-processing, and aggregation methods. Dashed red line indicates the 75th percentile cutoff score for funding designation. Shaded portions and labeled error bars represent the predicted amount of model sensitivity at a given percentile, in terms of how many percentile-ranks a tract can vary (e.g., in 95% of predictions, tracts at the 75th percentile can vary their score by 44 percentile-ranks).

Receiving algorithmic designation is financially consequential. We estimated through a causal analysis that the effect of receiving designation from the algorithm is a 104% (95% confidence interval: 62-145%) increase in funding, equivalent to $2.08 billion ($1.56-2.41 billion) in additional funding over a four-year period for 2,007 tracts (Fig S1, Table S2). Similarly, among the 400 tracts that would be eligible for designation under an alternative model (described below), we estimated they would have received equivalent to $632 million ($377-881 million) in additional funding over the same time period.

**Allocative tradeoffs and harm**

Under such a model with high uncertainty, every subjective model decision is implicitly a value judgment: who benefits from one particular version of a model, and who benefits from another? Both the model in its current form and plausible alternative forms may exhibit bias among different subgroups, illustrating the zero-sum nature of delegating funding allocation to a single model.

To exemplify these challenging tradeoffs, we constructed an alternative model for designation assignment. We (a) changed the pre-processing and aggregation methods to avoid penalizing tracts with extreme levels in variables such as air pollution indicators; and (b) incorporated a number of additional population health metrics for a more broad definition of vulnerability to environmental exposures. On average, incorporating these changes led to increased designation to tracts with higher levels of people of color in poverty, but decreased designation among populations of people of color overall (Fig. 2, S2).

In particular, expanding the "sensitive populations" category of the algorithm presents ethical concerns. The category is represented by three variables: respiratory health, cardiovascular health, and low birthweight. It would be sensible to include additional health indicators relevant to environmental exposures, such as chronic kidney disease or cancer (*10*, *11*). The inclusion of such variables, however, would result in the loss of designation for tracts with high Black populations. Because low birthweight disproportionately affects Black infants, the introduction of other variables such as cancer – which also disproportionately affects Black populations albeit to a lesser extent – would reduce the impact of low birthweight on the algorithm's output.

Moreover, we found the existing model to potentially underrepresent foreign-born populations. The model measures respiratory health in terms of emergency room visits for asthma attacks, which underrepresents groups who use the emergency room less or come from countries where asthma is less prevalent, yet still have other respiratory issues (*12*, *13*). Consequently, we found that using survey data of chronic obstructive pulmonary disease to represent respiratory health increases designation of tracts with foreign-born populations of 30% or higher (Fig. S3-4).

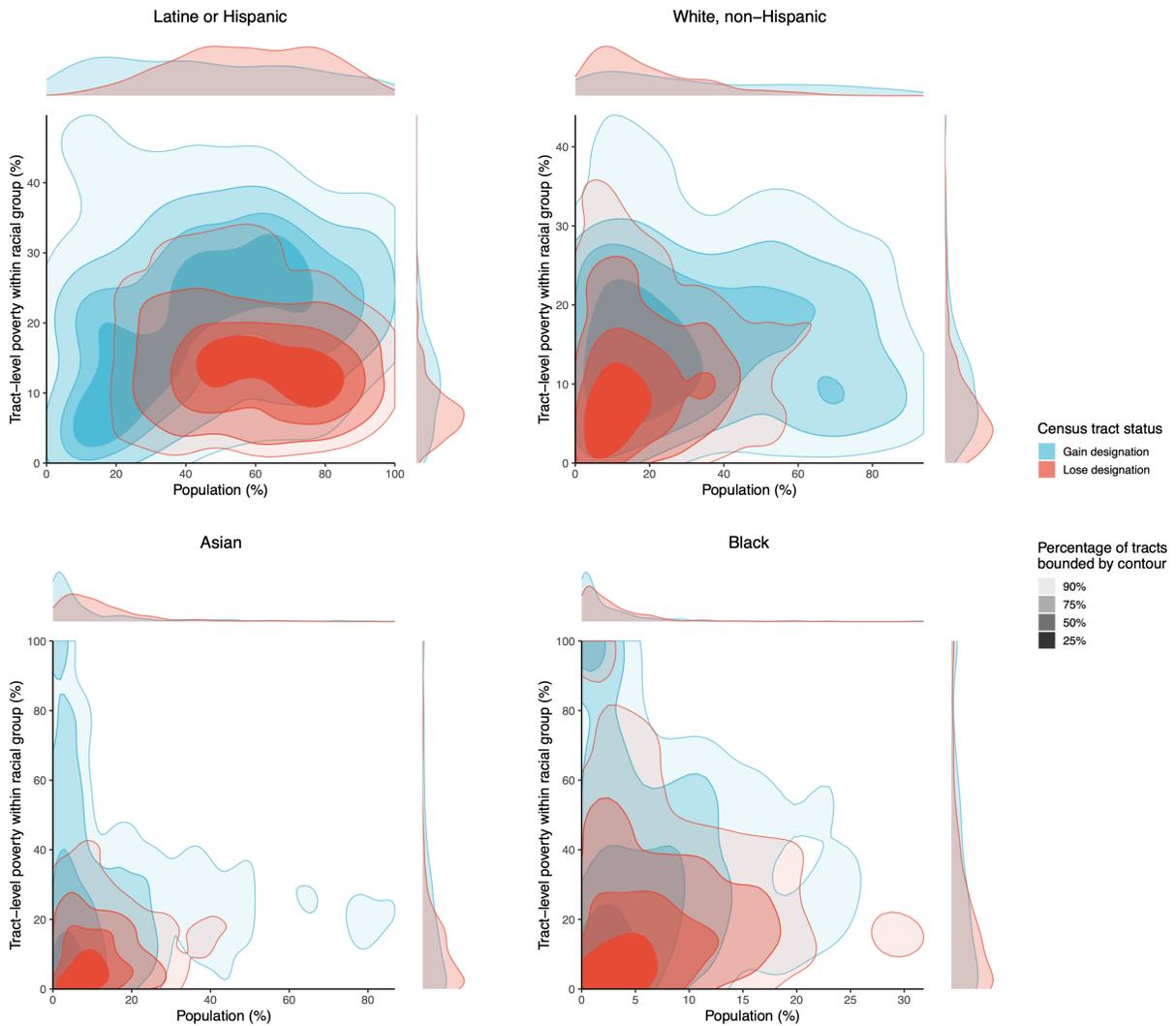

**Fig. 2. Allocative tradeoffs between populations of color in poverty and populations of color overall.** Comparison of how algorithmically designated tracts are distributed by race and poverty across the current CalEnviroScreen model and an alternative model, among tracts that would change designation status under the alternative model. The alternative model uses a different pre-processing technique, different aggregation technique, and incorporates additional population health variables. Red densities indicate tracts that receive designation under the current model but are not designated under the alternative model. Blue densities indicate tracts gaining designation under the alternative model. Contours are calculated as the smallest regions that bound a given proportion of the data.

The zero-sum nature and high sensitivity of the model are conducive to model manipulability. It is feasible for a politically motivated internal actor, whether subconsciously or intentionally, to prefer model specifications that designate tracts according to a specific demographic, such as political affiliation or race. Through adversarial optimization – optimizing over pre-processing and aggregation methods, health metrics, and variable weights – we find the maximum increase or decrease a model can be manipulated to favor a specific U.S. political party is 39% and 34%, respectively (Fig S5-6). Any efforts to mitigate the harms of allocative algorithms such as CalEnviroScreen thus need to consider both model sensitivity and manipulability.

**Mitigation strategies**

We propose assessing robustness of models through examining their sensitivity to alternative specifications, and incorporating additional models accordingly. For example, CalEPA recently decided to honor designations from both the current and previous versions of CalEnviroScreen, effectively taking the union of two different models. This approach reduces model sensitivity by 40.7%, and a three-model approach additionally incorporating designations from our alternative model reduces model sensitivity by 71.0%. Using multiple models also mitigates allocative harm: by broadening the category of who is considered disadvantaged, different populations are less likely to be in competition with each other for designation.

A potential concern is that increasing the number of designated tracts may dilute earmarked funds for disadvantaged groups. However, incorporating an additional model per our example would only increase the number of tracts by 10%, yet reduce model sensitivity by 51.1%. Doing so would also reduce equity concerns and more accurately represent the uncertainty inherent to designating tracts – consideration should be given as to whether these benefits outweigh the downsides. Furthermore, adding models is only one possible solution; other valid approaches that incorporate model uncertainty could include ensembling models or using tiered funding or lottery systems weighted by scores and uncertainty.

However, reducing model sensitivity is not a complete solution – transparency and accountability are necessary to reduce harm. The agency developing CalEnviroscreen is active in offering methodological transparency and soliciting feedback, which enables critiques such as ours and promotes public discourse – agencies developing similar tools to identify disadvantaged neighborhoods should follow suit. A safeguard like an external advisory committee comprising domain experts and leaders of local community groups could also help reduce harm by identifying ethical concerns that may have been missed internally. Doing so would also promote equitable representation and involvement from the public, aligning with the tool's goal of advancing environmental justice.

**Discussion**

Our findings are threefold: (1) CalEnviroScreen's model is both sensitive to change and financially consequential; (2) subjective model decisions lead to allocative tradeoffs, and models can be manipulated accordingly; and (3) model sensitivity can be mitigated by accounting for uncertainty in designations. Concretely, we recommend accounting for uncertainty by incorporating sensitivity analyses and potentially including additional models to increase model robustness, and urge for community-based independent oversight.

Our analysis is not a comprehensive audit of the tool. We do not identify every potential flaw or ethical concern of the model, but instead highlight illustrative examples of how model choices can facilitate allocative harm – only members of a given community can fully know how their respective tracts are represented and affected by the algorithm. Our estimates of model sensitivity are likely underestimates as we do not exhaustively specify alternative models. Further, our estimate of the funding impact of algorithmic designation is likely an underestimate, as detailed data on relevant private funding sources are not publicly available. Other limitations of our work are largely the same limitations besetting the data tool itself: key missing indicators such as indoor air pollution, a lack of measurement error metrics, and varying levels of data quality.

Technical and regulatory solutions will be necessary to address concerns of allocative harm as algorithms continue to be adopted for policy use. While misuse of such tools could exacerbate existing inequities, a careful and community-minded approach can lead to the broad realization of CalEnviroScreen's intended goal: furthering environmental justice and mitigating the harms done to structurally marginalized populations.

**Acknowledgments:** We thank Angela Yip and Erica Knox for helpful discussions. We thank Qiwei Lin, Kate Jen Li, and Feona Dong for research assistance with the Census block group analysis. We thank CalEPA, The Office of Environmental Health Hazard Assessment, and California Climate Investments for their transparency in methods and data, enabling this work.

**Funding:** BQH acknowledges support by the National Science Foundation Graduate Research Fellowship under Grant No. DGE 1656518. The contents of this article are solely the responsibility of the authors and do not necessarily represent the official views of any agency. Funding sources had no role in the writing of this manuscript or the decision to submit for publication.

**Authors' contributions:** BQH and ETC conceived of the research idea and designed and conducted the primary analyses. BQH drafted the manuscript. AK provided technical guidance regarding algorithmic fairness. DO performed the census block group analysis. DEH provided technical guidance regarding policy considerations. MVK and DHR supervised the work. All authors approved and contributed to the final version of the manuscript.

**Competing interests:** Authors declare that they have no competing interests.

**Data and materials availability:** All data used in this work are publicly available online. The CalEnviroScreen data can be found at https://oehha.ca.gov/calenviroscreen. The CDC PLACES data can be found at https://www.cdc.gov/places/index.html. The Climate Change Investments funding dataset can be found at https://www.caclimateinvestments.ca.gov/cci-data-dashboard. American Community Survey data can be found at https://www.census.gov/programs-surveys/acs/data.html. All code written for this work is available at https://github.com/etchin/allocativeharm, and all datasets used are archived at https://doi.org/10.7910/DVN/EVWNC2.

**Supplementary Materials**

Materials and Methods

Supplementary Text

Figs. S1 to S11

Tables S1 to S5

References (*15–18*)

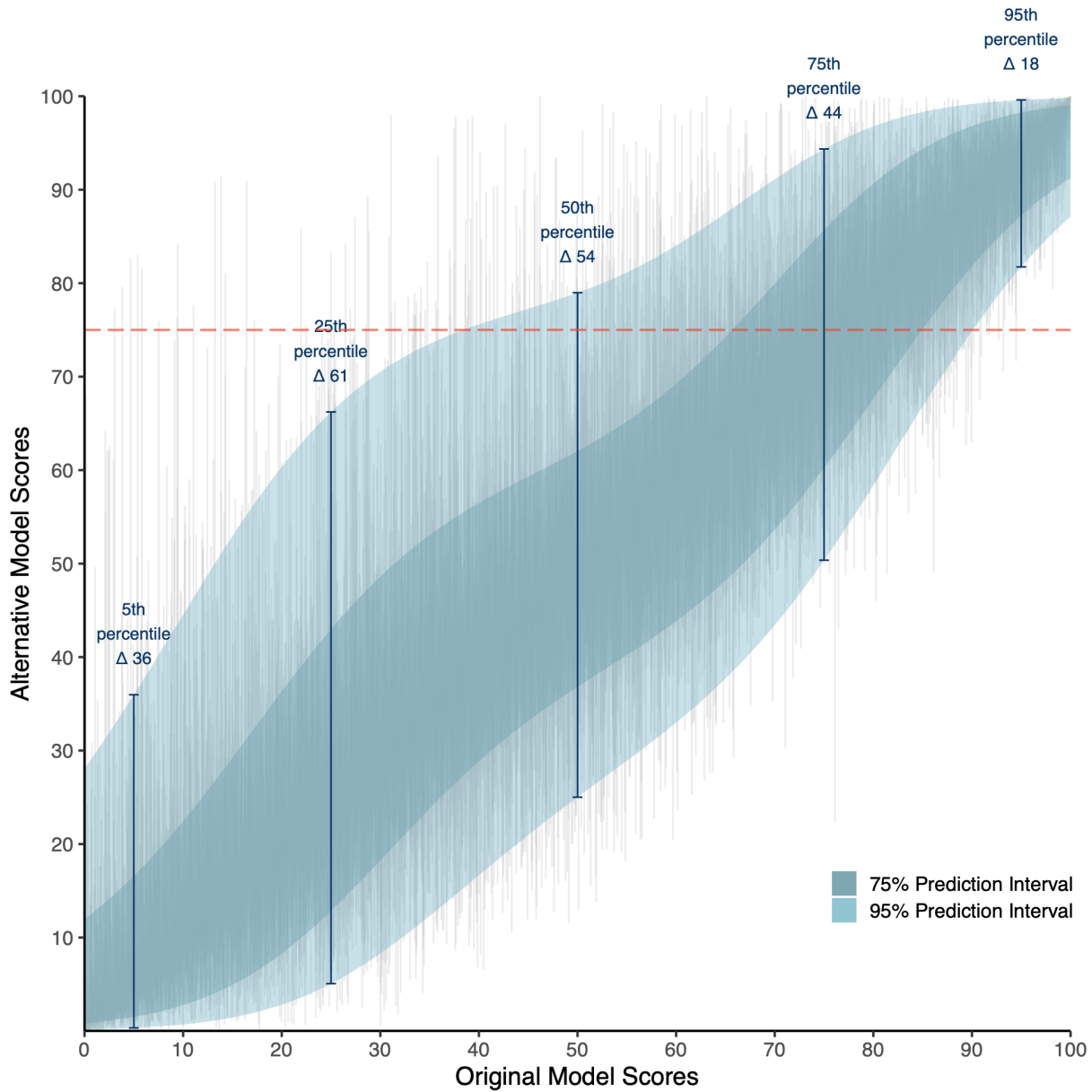

**Fig. 1. CalEnviroScreen's sensitivity to input parameters**. Axes denote model scores in terms of percentiles. Gray bars indicate maximum and minimum values from alternative plausible model specifications varying health metrics, pre-processing, and aggregation methods. Dashed red line indicates the 75th percentile cutoff score for funding designation. Shaded portions and labeled error bars represent the predicted amount of model sensitivity at a given percentile, in terms of how many percentile-ranks a tract can vary (e.g., in 95% of predictions, tracts at the 75th percentile can vary their score by 44 percentile-ranks).

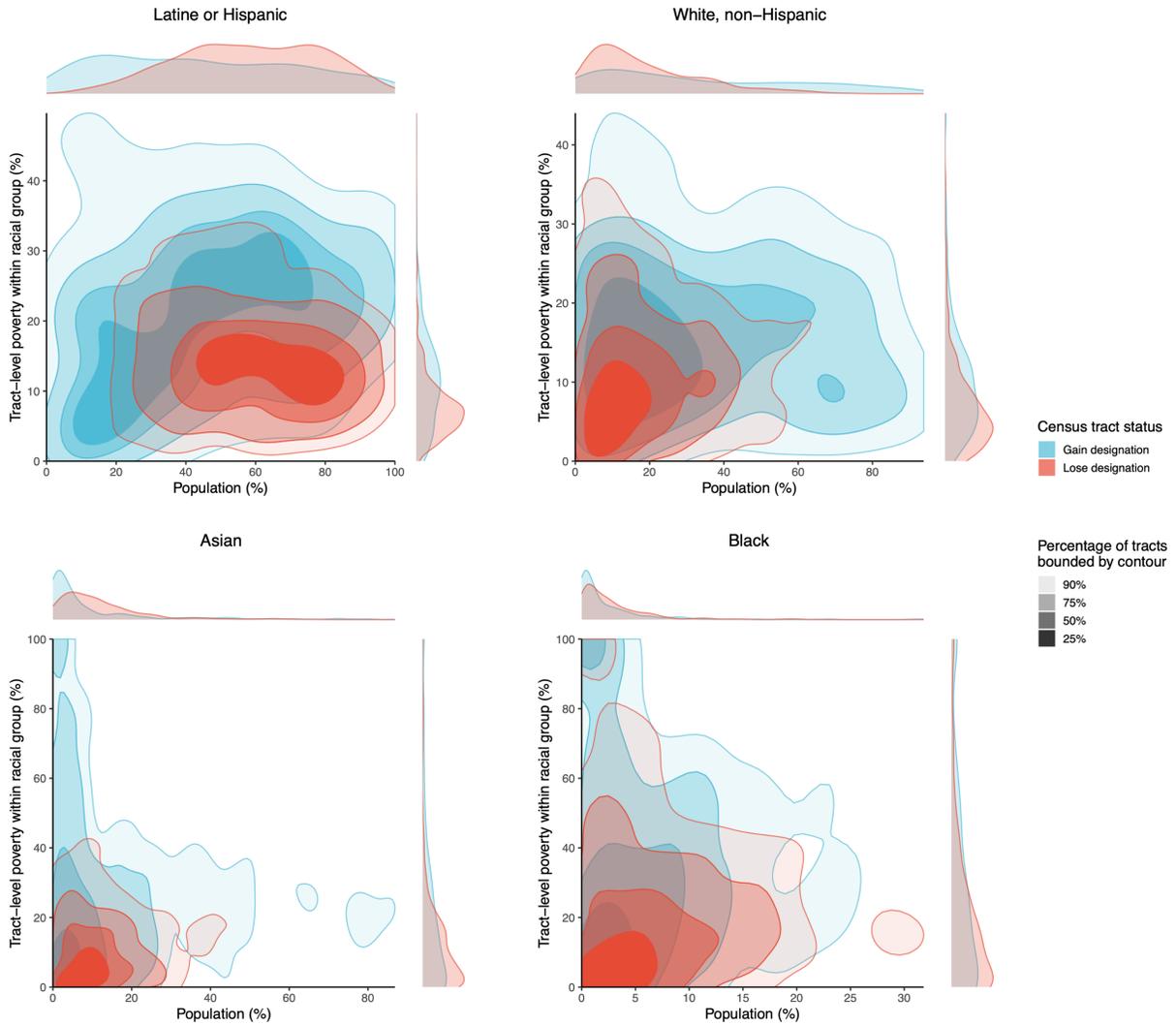

**Fig. 2. Allocative tradeoffs between populations of color in poverty and populations of color overall.** Comparison of how algorithmically designated tracts are distributed by race and poverty across the current CalEnviroScreen model and an alternative model, among tracts that would change designation status under the alternative model. The alternative model uses a different pre-processing technique, different aggregation technique, and incorporates additional population health variables. Red densities indicate tracts that receive designation under the current model but are not designated under the alternative model. Blue densities indicate tracts gaining designation under the alternative model. Contours are calculated as the smallest regions that bound a given proportion of the data.

# Supplementary Materials for

## Potential for allocative harm in an environmental justice data tool

Benjamin Q. Huynh, Elizabeth T. Chin, Allison Koenecke, Derek Ouyang, Daniel E. Ho,
Mathew V. Kiang, David H. Rehkopf

Corresponding author: benhuynh@stanford.edu

**The PDF file includes:**

Materials and Methods
Supplementary Text
Figs. S1 to S11
Tables S1 to S5
References 15-18

**Materials and methods**

Data

For our sensitivity analyses, we used census tract-level data obtained from the current version of CalEnviroScreen (version 4.0, implemented in late 2021), consisting of 8,035 observations with 21 variables measuring different aspects of environmental exposures and population characteristics. The variables measure ozone levels, fine particulate matter, diesel particulate matter, drinking water contaminants, lead exposure, pesticide use, toxic release from facilities, traffic impacts, cleanup sites, groundwater threats, hazardous waste, impaired waters, solid waste sites, asthma, cardiovascular disease, low birth weight, education, housing burden, linguistic isolation, poverty, and unemployment.

For our additional variable analyses, we used the PLACES dataset from the Centers for Disease Control and Prevention to include tract-level variables on estimated prevalence for asthma, cancer, chronic kidney disease, and coronary heart disease. We obtained demographic information on tract-level race/ethnicity from the American Community Survey (ACS).

For the causal analysis specifically, we examined years 2017-2021 of the California Climate Investments (CCI) funding dataset, and used CalEnviroScreen 3.0 scores (implemented in 2017), as there is not yet sufficient data for funded projects guided by CalEnviroScreen 4.0. We calculated the amount of funding allocated to each census tract by summing the amount of funding received from different programs for each tract. For funding projects that were attributed to assembly districts but not specific census tracts, we attributed them to tracts using the following steps: (1) funds earmarked for "priority populations" such as disadvantaged tracts were exclusively attributed to their respective tracts within that district; (2) remaining funds were attributed to non-priority population tracts within the district up to the amount attributed to priority population tracts; (3) and any remaining funds after that were distributed equally (see Supplement for more details). To attribute districts to tracts spanning multiple districts, we followed the methodology listed in the CCI funding dataset: we solely considered them to belong to whichever district contained the largest population. For tracts that were missing relevant block-level population metrics, we assigned them to districts based on whichever district contained more blocks from the given tract. For the single tract that had missing population metrics and the same number of blocks for two districts, we assigned its district based on geographical area.

Sensitivity Analyses

We first reproduced the original CalEnviroScreen model based on its documentation, and validated our reproduction on existing data. We then identified potential issues in the data tool and conceived plausible alternative models. As a general approach, we built alternative models (implementing various small changes to the current CalEnviroScreen model) and evaluated how they differed from the original model to assess sensitivity of the CalEnviroScreen algorithm to model decisions. Variation was measured in terms of percent change in tracts changing designation, i.e., the number of tracts changing designation divided by the total number of tracts times 100. Details of each step of this approach are below.

We assessed changes to (a) pre-processing, (b) aggregation methods, and (c) health metrics, all subjective areas for constructing composite indicators. We assessed pre-processing methods by changing the existing pre-processing method, percentile-ranking, to $z$-score standardization. We assessed aggregation methods by changing the existing aggregation method, multiplication, to arithmetic mean.

We assessed health metrics based on our concerns of public health biases perpetuated by the algorithm. First, we noted that the existing method of measuring health indicators strictly by emergency room visits may be skewed towards populations who use the ER disproportionately often, and so tested including tract-level survey indicators of health in the model, namely asthma and cardiovascular health (*13*). Secondly, we noted that only using asthma as a measure of environmental vulnerability with respect to respiratory health may not be fully reflective of those with respiratory health issues, so we tested including survey indicators for chronic obstructive pulmonary disease. The inclusion of survey indicators of health were weighted such that categories of respiratory health, cardiovascular health, and low birth-weight were equally weighted. Lastly, we noted that low birth-weight, cardiovascular, and respiratory issues are not the only health-related ways in which populations may be vulnerable to environmental exposures, and so tested including other indicators, such as chronic kidney disease and cancer.

Our alternative model was pre-specified and designed based on the changes listed above: changing pre-processing to standardization, changing aggregation to averaging, and including survey indicators of health for cardiovascular health, asthma, COPD, chronic kidney disease, and cancer. We calculated the overall model sensitivity by assessing the different combinations of model specifications by varying pre-processing, aggregation, and health variables – $z$-score standardization vs. percentile ranking, multiplication vs. averaging, and including vs. excluding the additional health variables we specified – and calculating the number of distinct tracts that change designation across all models. We trained a smooth nonparametric additive quantile regression model on the range (i.e., minimum and maximum values across models) for each tract to obtain prediction intervals (*14*).

Empirical Strategy

We used a sharp regression discontinuity design with local linear regression as the functional form to estimate the effect of algorithm designation on total funding received (*15*). We selected the bandwidth using the Imbens-Kalyanaraman algorithm (*16*). The treatment variable was a binary indicator for each tract denoting whether it was designated as disadvantaged by the algorithm. The outcome variable was the log of total funding received per tract. The forcing variable was the CalEnviroScreen percentile rank for each tract. Covariates included the aggregate pollution burden and population characteristics indicators from CalEnviroScreen, and tract-level race and poverty estimates from ACS. As robustness checks, we estimated the treatment effect with varying bandwidths, functional forms, covariate adjustments, and dataset configurations. We also estimated the treatment effect with a propensity score matching approach, and two machine learning approaches: a linear model forest and a doubly-robust targeted learning approach, and (see Supplement). All parenthetical values reported in the main text are 95% confidence intervals, and were calculated by multiplying standard errors by the 97.5th percentile point of the standard normal distribution.

Adversarial Optimization

We formulated our optimization strategy as follows:

$$\max_{W,p,a} \quad \phi_d(f(W,p,a))$$
$$\text{s.t.} \quad 0.1 \leq w_i \leq 0.9$$
$$p \in \{0,1\}$$
$$a \in \{0,1\}$$

where *f* is the CalEnviroScreen algorithm designating tracts as disadvantaged, $\Phi_d$ is a function totaling the number of tracts belonging to a chosen demographic *d* (e.g., political affiliation, race), $W = \{w_1,...,w_n\}$ is a vector of weights for each variable in the CalEnviroScreen algorithm, *p* is an indicator variable denoting pre-processing options (percentile-ranking vs. *z*-score standardization), and *a* is an indicator variable denoting aggregation methods (multiplication vs. averaging). Weight variables were restricted to be between 0.1 and 0.9 to prevent extreme individual weight values. We used the Hooke-Jeeves method to solve the optimization problem (*17*).

Political affiliation at the tract-level was determined by party affiliation in terms of assembly district. For tracts that spanned multiple assembly districts, we attributed those tracts to the districts in which most of their population belonged, in line with how the Climate Change Investments fund attributes tract-level funding to tracts spanning multiple districts. Race was determined by percentage of the population for each tract being of a given race. We calculated the percent change in designated tracts for the party with fewer tracts.

**Supplementary Text**

CalEnviroScreen background

In 2012, California Senate Bill 535 was signed into law, allocating 25% of California's proceeds from its cap-and-trade program to social welfare funds for environmentally disadvantaged communities. California's Office of Environmental Health Hazard Assessment (OEHHA) developed CalEnviroscreen as part of the California Environmental Protection Agency's environmental justice program, and is used to designate disadvantaged communities as mandated by SB 535. In 2016, California Assembly Bill 1550 amended SB 535 to require GGRF funds that benefit disadvantaged communities to be located within those communities. AB 1550 also earmarked an additional 10% of funds to be dedicated to low-income communities, and half of those funds are reserved for low-income households and census tracts in "buffer regions", areas within a half mile of a designated disadvantaged tract (Fig. S7).

The current version is CalEnviroScreen 4.0 – the model is updated periodically to use more recent variables, update methodology, and make changes according to feedback. Public workshops are made available to discuss the methodology of CalEnviroScreen, and when new versions are being actively developed, public comment periods are held to solicit feedback. The process of updating CalEnviroScreen is reported to be separate from the process of designating disadvantaged communities; despite the fact that OEHHA is housed within CalEPA, OEHHA is responsible for developing CalEnviroScreen, and CalEPA is responsible for designating disadvantaged communities, for which they have historically used CalEnviroScreen as the primary basis.

It is for this reason that the disadvantaged communities designation is slightly broader than just the top 25% of scores (1984 tracts) from CalEnviroScreen: the disadvantaged communities designation was expanded in 2022 to include lands under federally recognized Tribes, census tracts lacking overall scores but receiving the highest 5% of pollution burden scores (19 tracts), and tracts that received designation under the previous version of CalEnviroScreen (307 tracts). Our analysis primarily focuses on the methodology of CalEnviroScreen and does not extend to 2022, so we largely do not consider these additional disadvantaged communities, except to assess the impact of including the previous version of CalEnviroScreen on model sensitivity.

CalEnviroScreen takes Census tract-level data on pollution burden and population characteristics and reduces it into a single score; the Census tracts in the top 25% of scores are designated "disadvantaged communities." The pollution burden category is split into two subcategories: exposures and environmental effects; the population characteristics category is also split into two subcategories: socioeconomic factors and sensitive populations. All categories are weighted equally except for environmental effects, which is half-weighted. Variables in each subcategory are converted to ranked percentiles then averaged together into a subcategory score. The subcategory scores are then averaged into category scores (pollution burden and population characteristics), which are then multiplied together to create a CalEnviroScreen score. Our implementation of CalEnviroScreen's model was reproduced according to their written methodology – in our validation comparing our scores to theirs, we found our implementation had 100% accuracy in matching their designations, and that our implementation's scores were accurate up to eight decimal points. Differences beyond that are likely due to floating point imprecision.

Rationale for pre-processing and aggregation measures

We opted to test *z*-score standardization as a pre-processing method instead of CalEnviroScreen's percentile ranking approach because it is a widely-accepted method likely more appropriate for the data (*9*). Percentile ranking converts the original distributions of the raw data to uniform distributions: extreme values become underweighted and values near the center of the distribution become more highly weighted. For example, the median difference across variables between the 60th and 80th percentiles is relatively low (0.6 standard deviations), whereas the median difference between the 80th and 99th percentiles is high (2.6 standard deviations). Pre-processing through percentile ranking erroneously equates these two differences.

Next, we tested using additive aggregation over multiplicative aggregation because it is similarly a commonly used method likely more appropriate for the data (*9*). Multiplication is less compensatory than additive aggregation: under multiplicative aggregation, low scores in one category cannot be fully compensated for by high scores in another category, thus necessitating above-average values in both categories for a high final score (*9*). For example, a tract with high pollution burden and population characteristic scores of (9.2, 4.3) does not make the cutoff, yet a tract with moderate scores of (6.5, 6.1) does. Considering the small differences between scores closer to the center of the distribution, additive aggregation is potentially more appropriate for the data.

Model sensitivity from omitting variable categories

To identify the different types of projects for which funding is allocated, we manually categorized each funding project in the CCI funding dataset into six different categories: air quality; climate adaptation and infrastructure; food and agriculture; housing, employment, and community development; transit; and utilities (Fig. S8). Because the funding earmarked for disadvantaged communities is implemented across a broad variety of sectors, we hypothesized a single index may not be appropriate for addressing the diverse challenges and development opportunities faced by disadvantaged communities. For example, among the four variable categories of the algorithm, pollution may not be the most informative category for determining where to allocate affordable housing or agricultural subsidies.

We assessed the sensitivity of the model to removing one of the four variable categories (environmental exposures, environmental effects, sensitive populations, and socioeconomic factors). We found that 7.4-16.9% of census tracts change designation upon removing a variable category (Table S3), supporting our hypothesis that consideration may be needed as to whether a single composite indicator is appropriate for allocating funding across different sectors.

Variable weighting and rationale for adversarial optimization

We measured the correlations between variables in CalEnviroScreen, as well as their individual ability to predict whether a tract would be designated as a disadvantaged community. We found that many variables in CalEnviroScreen's model are strongly correlated with each other, and that each variable's ability to predict designation is almost entirely explained by its level of correlation to the other variables ($R^2 = 0.96$). Highly correlated variables like education levels can single-handedly predict designation to an accurate degree (AUC = 0.91). Conversely, environmental variables like pesticide exposure (AUC = 0.52) or impaired water bodies (AUC = 0.53) have negligible predictive power (Figs. S9-10).

We thus chose to include variable weighting in our adversarial optimization experiments for two reasons: (1) there is precedent for it, in that all of the variables in the environmental effects group are down-weighted by half; and (2) the existing weighting scheme where all other variables are equally weighted is in itself a subjective design choice, with the impact of implicitly granting highly correlated variables more importance in the model.

Census block group analysis

We measured the effect of using higher resolution data for socioeconomic indicators on overall CalEnviroScreen output. For all socioeconomic variables except for housing burden, we switched the data from tract-level to census block group (CBG) level using ACS data. For the housing burden variable, there is no publicly available CBG-level dataset that exactly measures the one used by CalEnviroScreen, so we created a close approximation using available ACS data by separately counting renters and housing-burdened owners.

For renters, we used the Household Income by Gross Rent as a Percentage of Income measure from ACS, identified tiers for each county containing the 80% area median income definition and counted all renters up to and including that income tier, effectively identifying the proportion of renters in this group with housing burden over 50%. For homeowners, we used the Age of Householder by Selected Monthly Owner Costs as a Percentage of Household Income measure from ACS, and estimated the number of housing-burdened households in each income tie using an equal distribution assumption between age-of-household and income. We followed the rest of the procedure as for the renters. In this group, we measure housing burden over 35% instead of 50% due to limited data availability.

Combining both the renters and homeowners estimates for housing burden, we thus construct a variable that estimates over 35% or 50% housing burden for all households at the CBG-level. As validation, we find this approach at the tract-level has a 0.95 Pearson correlation with the existing CalEnviroScreen housing burden variable. To convert our CBG-level data back to tract-level data, we calculate the weighted mean of CBG variable values, weighted by CBG population, then convert to percentile ranks to reconstruct the socioeconomic factors portion of the CalEnviroScreen method. We find incorporating this change leads to 1.9% of tracts changing designation, and incorporating it into our main sensitivity analysis leads to 16.5% of tracts changing designation.

Attributing district-level earmarked funding to census tracts

For funding projects that were listed at the district level, but did not specify individual census tracts, we attributed their earmarked funding (i.e., disadvantaged community, low-income household/tract, buffer region tracts) to their respective census tracts within the district. Any remaining funding was attributed to non-priority populations up to the same amount attributed to priority-populations. If any funding remained after that, funding was attributed equally among tracts within the district.

The CCI dataset was subject to a number of errors for which we attempted to account. One funding project (out of 4203) was listed with a negative funding amount and was removed prior to analysis. 158 projects had listed funding totals that were less than the sum of their earmarked funding, indicating potential errors in data entry: we assumed the funding totals to be correct because their sum aligned with the amount listed on the CCI website, and set the relevant earmarked category to be equal to the funding total; if multiple earmarked categories had equal amounts of funding, funding was attributed to be evenly split between them; if there were nonzero equal values across earmarked categories, funding was divided from the total sum according to the ratio of funding values between earmarked categories.

Empirical strategy rationale and robustness checks

We selected our empirical strategy to estimate the effect of receiving designation on overall funding received – using a sharp regression discontinuity design with local linear regression as the functional form – to exploit the clear discontinuity at the 75th percentile sharp threshold of CalEnviroScreen scores, where the treatment assignment mechanism is fully known (Fig. S11).

We opted for a local linear regression functional form for its rate optimality and to avoid noisy implicit weights often resulting from higher-order functional forms (*16, 18*).

Formally, we specify our model as follows:

$$Y_i = \alpha + \tau D_i + \beta f(X_i) + \gamma f(X_i) D_i + Z_i \delta + \epsilon_i$$

where $Y_i$ is the log of total funding for census tract $i$, α is the intercept, $D_i$ is an indicator variable denoting whether tract $i$ received designation as a disadvantaged tract, $X_i$ is the CalEnviroScreen percentile rank for tract $i$, $Z_i$ represents other tract-level covariates used for adjustment, and $f(X_i)$ represents the functional form used to model the relationship between CalEnviroScreen percentile rank and total funding.

One of our primary findings in the paper was that receiving algorithmic designation results in a 104% increase in funding. We performed a number of robustness checks to test the sensitivity of this result to different model specifications. We varied the bandwidth (Imbens-Kalyanaraman optimal bandwidth vs 10), functional forms (local linear vs quadratic), and level of covariate adjustment (adjusting for race, poverty, and the two primary CalEnviroScreen indicators: Pollution Burden and Population Characteristics, vs no covariate adjustment), and found all estimates to have overlapping confidence intervals with our primary model specification. We repeated the above robustness checks across different dataset specifications: (1) the main specification of including all projects funded from 2017-2021, (2) an alternative specification of only including projects with funding specifically attributed to CalEnviroScreen 3.0 (as opposed to previous versions of CalEnviroScreen), and (3) an alternative specification of including all projects from 2017-2021 except for projects related to high-speed rail installations, which are funding projects technically attributed to specific census tracts but benefitting populations far broader than the area in which they are built. Across the dataset specifications, we also found all estimates to have overlapping confidence intervals with our primary model specification (Table S2).

As another robustness check, we also estimated the causal effect of algorithmic designation on funding using a linear model forest in a sharp regression discontinuity design (*15*). We used the same specifications as our main regression discontinuity analysis, except with more covariates: all variables from CalEnviroScreen (except for the treatment/running variables), and demographic variables on race and poverty. We used the *grf* package in R 3.6.3, using 10,000 trees per forest, specified a triangular kernel, and grew 20 trees per subsample to calculate confidence intervals. We varied over bandwidths and dataset specifications as mentioned above, finding all estimates to have overlapping confidence intervals with our primary model specification's estimate of a 104% increase in funding (Table S4).

Matching based secondary analysis

As a secondary causal analysis, we also employed a propensity score matching technique to estimate the effect of algorithmic designation on overall funding. Because matching approaches require complete datasets, we employed imputation methods to account for missing data. Among all census tracts with CalEnviroScreen scores, 4.7% of tracts have at least one missing value. Eight of the pollution and population variables used in the CalEnviroScreen algorithm contain at least one missing value, of which the maximum missingness in any variable was 1.8%. While the CalEnviroScreen algorithm handles missingness by calculating the score over all non-missing variables, we use multiple imputation to handle missingness for sensitivity analyses of matching methods for causal inference.

We used the *mice* package (version 3.15.0) in R to generate 10 imputed datasets by predictive mean matching, with a maximum iteration of 50. In addition to the pollution and population variables used in the CalEnviroScreen algorithm, we included location (county) and demographic (race) information to facilitate the imputation process. After imputation, we conducted matching on each imputed dataset separately then pooled the results together.

We employed matching as an orthogonal causal inference approach to assess sensitivity of estimates. We used the *MatchIt* package (version 4.5.2) in R to match each DAC census tract with a non-DAC tract according to the raw data from pollution and population variables used in the CalEnviroScreen algorithm. We employed nearest neighbor matching on each imputed dataset separately. Standardized mean differences were calculated on the matched data, and any covariates with a standardized mean difference greater than 0.2 were included as covariate adjustments in the outcome model.

Calipers were used to ensure that the distance between matched units were within a specific caliper width, and any units that did not have available matches within the caliper were dropped from the matched sample. We varied over several calipers (0.1, 0.2, and 0.3) in the sensitivity analyses, with calipers measured in standard deviation units. We also varied over three propensity score algorithms: generalized linear models, generalized additive models, and covariate balancing propensity score.

To calculate the pooled means for each matching specification, we used the following formula:

$$\bar{\beta} = \frac{1}{m} \sum_{i=1}^{m} \beta_i$$

where $\bar{\beta}$ is the pooled parameter estimate, $m$ is the number of imputed datasets, and $\beta_i$ is the parameter estimate for each imputed dataset $i$.

The pooled standard error can be decomposed into the variances within and between imputed datasets. Within imputation variance represents the average sample variance estimated in each imputed dataset.

$$Var_W = \frac{1}{m}\sum_{i=1}^{m} SE_i^2$$

where $Var_W$ is the within imputation variance and $SE_i$ is the standard error estimated in each imputation dataset $i$.

Between imputation variance reflects variance from missingness. We estimate this by taking the variance of the estimand over imputed datasets.

$$Var_B = \frac{1}{m-1}\sum_{i=1}^{m}(\beta_i - \bar{\beta})^2$$

where $Var_B$ is the between imputation variance.

The pooled variance and standard errors were then estimated using the following formulas:

$$Var_{Pooled} = Var_W + Var_B + \frac{1}{m}Var_B$$

$$SE_{Pooled} = \sqrt{Var_{Pooled}}$$

where $Var_{Pooled}$ is the pooled variance and $SE_{Pooled}$ is the pooled standard error.

**Table S1: Model sensitivity measured as percentage of total tracts that change designation under different model specifications.** "Scaling" refers to using z-score standardization instead of percentile ranking for pre-processing variables. The "Additional health metrics" row refers to whether additional population health variables were incorporated. "Overall" refers to measuring model sensitivity when taking the union of designations from all possible model specifications across scaling, mean, and health.

| Pre-processing | Scaling | | | | Percentile-ranking | | | | |
|---|---|---|---|---|---|---|---|---|---|
| Aggregation | Additive | | Multiplicative | | Additive | | Multiplicative | | |
| Additional health metrics | All | None | All | None | All | None | All | None | Overall |
| % Change in designation | 9.8 | 8.4 | 7.9 | 5.4 | 4.8 | 1.4 | 4.3 | *Ref.* | 16.1 |

**Table S2: Table of effect size estimates from a sharp regression discontinuity approach with varying functional forms, bandwidths, and covariate adjustments.** Sensitivity analyses were conducted by varying over model specifications (functional form, bandwidth, and covariate adjustments), as well as across different specifications of the funding dataset: (A) All projects funded from 2017-2021; (B) All projects that explicitly note their funding decisions were based on CalEnviroScreen 3.0.; and (C) All projects funded from 2017-2021 except for projects related to the high-speed rail initiative. Point estimates represent the estimated percent increase in funding resulting from receiving designation. Values in parentheses represent 95% confidence intervals. IK refers to the optimal bandwidths computed using the Imbens-Kalyanaraman algorithm. "All" refers to adjusting for tract-level race, poverty, and CalEnviroScreen's two primary indicators of pollution burden and population characteristics. Bolded text represents the effect size of the main analysis.

**A.**

|  | Specification | | | | | | | |
|---|---|---|---|---|---|---|---|---|
| Functional form | Local Linear | | | | Quadratic | | | |
| Bandwidth | 3.86 (IK) | | 10 | | 3.86 (IK) | | 10 | |
| Covariate adjustment | All | None | All | None | All | None | All | None |
| Effect size estimate (95% CI) | **103.5 (61.9, 145.1)** | 103.8 (60.2, 147,3) | 117.2 (93.2, 141.2) | 104.2 (79.3, 129.2) | 115.8 (47.2, 184.4) | 106.8 (34.9, 178.8) | 101.6 (63.7, 139.6) | 92.7 (53.0, 132.4) |

**B.**

|  | Specification | | | | | | | |
|---|---|---|---|---|---|---|---|---|
| Functional form | Local Linear | | | | Quadratic | | | |
| Bandwidth | 3.62 (IK) | | 10 | | 3.62 (IK) | | 10 | |
| Covariate adjustment | All | None | All | None | All | None | All | None |
| Effect size estimate (95% CI) | 133.0 (94.2, 171.8) | 135.1 (94.9,175.2) | 167.9 (146.3, 189.4) | 152.7 (130.3, 175.0) | 155.3 (91.6, 219.0) | 140.9 (74.8, 207.0) | 125.9 (92.1, 159.7) | 116.9 (81.5, 152.2) |

C.

| | Specification | | | | | | | |
|---|---|---|---|---|---|---|---|---|
| Functional form | Local Linear | | | | Quadratic | | | |
| Bandwidth | 3.37 (IK) | | 10 | | 3.37 (IK) | | 10 | |
| Covariate adjustment | All | None | All | None | All | None | All | None |
| Effect size estimate (95% CI) | 100.5 (60.7-140.4) | 104.0 (63.0-145.0) | 122.0 (100.9-143.1) | 109.5 (87.6-131.4) | 99.7 (34.1-165.4) | 88.5 (21.0-156.0) | 93.6 (60.4-126.7) | 87.5 (52.9-122.1) |

**Table S3: Model sensitivity measured as percentage of total tracts that change designation when removing specific variable categories.**

| Omitted variable category | Environmental Exposure | Environmental Effects | Sensitive Populations | Socioeconomic factors |
|---|---|---|---|---|
| % Change in designation | 16.9 | 7.4 | 7.7 | 8.1 |

**Table S4: Table of effect size estimates from a sharp regression discontinuity approach using a linear model forest with varying bandwidths and dataset specifications.** Values in parentheses represent 95% confidence intervals. IK refers to the optimal bandwidths computed using the Imbens-Kalyanaraman algorithm.

| Dataset Specification | Funding projects from 2017-21 | | Funding projects attributed to CES 3.0 | | No high-speed rail projects | |
|---|---|---|---|---|---|---|
| Bandwidth | 3.86 (IK) | 10 | 3.62 (IK) | 10 | 3.37 (IK) | 10 |
| Effect size estimate | 113.9 (74.1-153.6) | 115.7 (56.2-175.2) | 151.6 (116.2-186.9) | 159.7 (97.2 -222.1) | 109.2 (72.4-146.0) | 115.2 (55.4-175.0) |

**Table S5: Table of effect size estimates from matching approaches with varying matching methods and calipers.** Sensitivity analyses were conducted by varying over model specifications (matching method, caliper), as well as across different specifications of the funding dataset: (A) All projects funded from 2017-2021; (B) All projects that explicitly note their funding decisions were based on CalEnviroScreen 3.0; and (C) All projects funded from 2017-2021 except for projects related to the high-speed rail initiative. Multiple imputation was used to handle missingness. Point estimates represent the mean estimated percent increase in funding resulting from receiving designation over 10 imputed datasets. Values in the parentheses represent 95% confidence intervals, with pooled standard errors reflecting within and between sampling variance of mean differences in imputed datasets. Sample size means and standard deviations are also reported.

**A.**

| Matching method | Generalized Linear Model | | | Generalized Additive Model | | | Covariate Balancing Propensity Score | | |
|---|---|---|---|---|---|---|---|---|---|
| Caliper | 0.1 | 0.2 | 0.3 | 0.1 | 0.2 | 0.3 | 0.1 | 0.2 | 0.3 |
| Effect size estimate (95% CI) | 92.8 (68.3, 120.8) | 91.1 (66.3, 119.7) | 83.2 (59.7, 110.2) | 92.7 (68.2, 120.7) | 91.1 (66.1, 119.8) | 83.2 (59.7, 110.1) | 84.8 (56.1, 118.8) | 87 (56.9, 122.8) | 89.4 (63, 120.2) |
| Mean sample size (sd) | 1151 (4.64) | 1213.8 (4.47) | 1283.2 (3.55) | 1151 (4.64) | 1213.8 (4.47) | 1283.2 (3.55) | 1269 (23.35) | 1339.8 (23.2) | 1406 (26.53) |

**B.**

| Matching method | Generalized Linear Model | | | Generalized Additive Model | | | Covariate Balancing Propensity Score | | |
|---|---|---|---|---|---|---|---|---|---|
| Caliper | 0.1 | 0.2 | 0.3 | 0.1 | 0.2 | 0.3 | 0.1 | 0.2 | 0.3 |
| Effect size estimate (95% CI) | 127.1 (100.7, 156.9) | 125 (99.1, 154.3) | 116.4 (91.4, 144.6) | 127 (100.6, 156.8) | 124.9 (99, 154.3) | 116.3 (91.3, 144.6) | 117.7 (86.6, 154.1) | 119.3 (88.6, 155) | 120.9 (93.8, 151.7) |
| Mean sample size (sd) | 1151 (4.64) | 1213.8 (4.47) | 1283.2 (3.55) | 1151 (4.64) | 1213.8 (4.47) | 1283.2 (3.55) | 1269 (23.35) | 1339.8 (23.2) | 1406 (26.53) |

**C.**

| Matching method | Generalized Linear Model | | | Generalized Additive Model | | | Covariate Balancing Propensity Score | | |
|---|---|---|---|---|---|---|---|---|---|
| Caliper | 0.1 | 0.2 | 0.3 | 0.1 | 0.2 | 0.3 | 0.1 | 0.2 | 0.3 |
| Effect size estimate (95% CI) | 87 (65.4, 111.4) | 86.4 (64.8, 110.8) | 79.1 (58.8, 101.9) | 86.9 (65.3, 111.3) | 86.3 (64.6, 110.8) | 79 (58.8, 101.9) | 79.6 (55.7, 107.2) | 81.8 (57.6, 109.6) | 83.4 (61.8, 107.9) |
| Mean sample size (sd) | 1151 (4.64) | 1213.8 (4.47) | 1283.2 (3.55) | 1151 (4.64) | 1213.8 (4.47) | 1283.2 (3.55) | 1269 (23.35) | 1339.8 (23.2) | 1406 (26.53) |

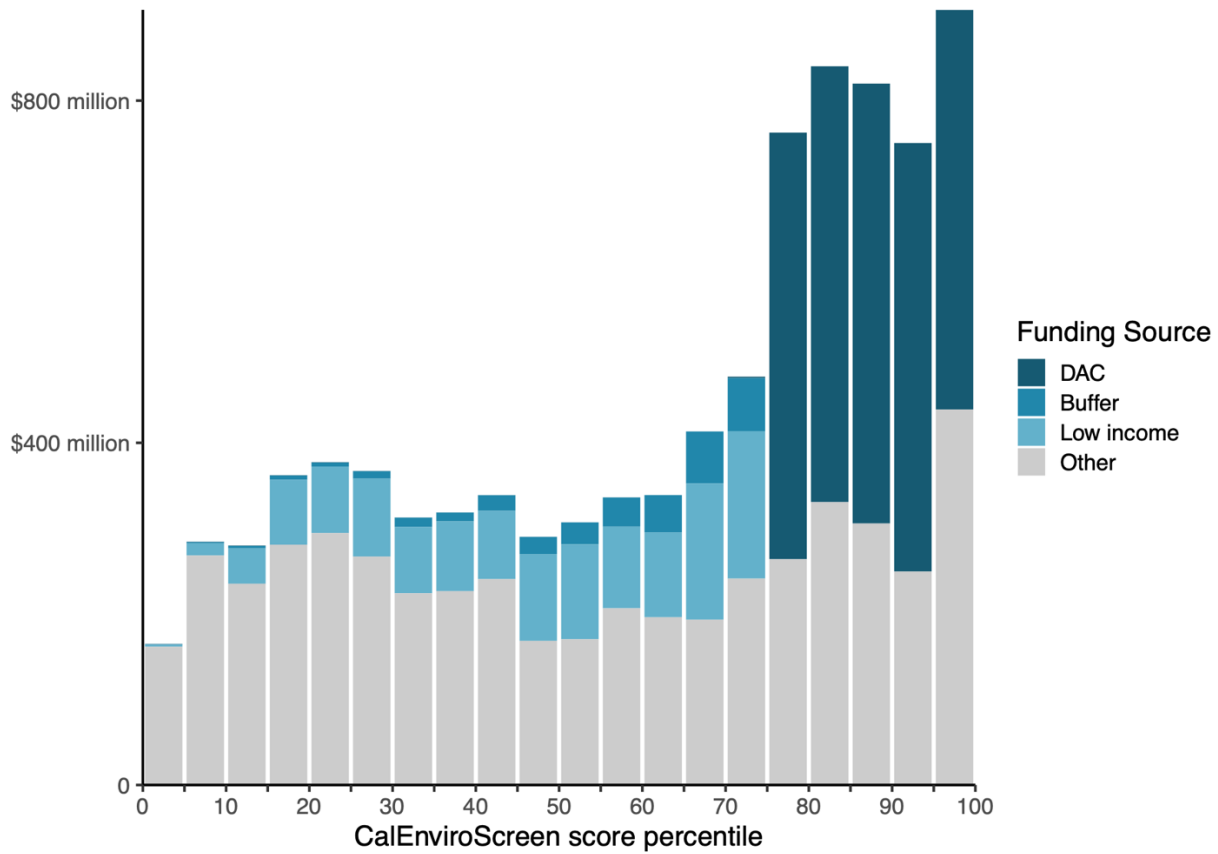

**Fig. S1. Total cumulative funding by California Climate Investments implemented in census tracts by CalEnviroScreen percentile from 2017-2021.** Dark blue bars indicate funding earmarked specifically for disadvantaged communities (DAC); Lighter blue bars indicate other earmarked funding (Buffer and Low income); Gray bars indicate all other funding. Buffer funding is earmarked for low-income communities and households which are not designated as DAC, but are within half a mile from a DAC census tract. Low income funding is earmarked for low-income communities and households statewide.

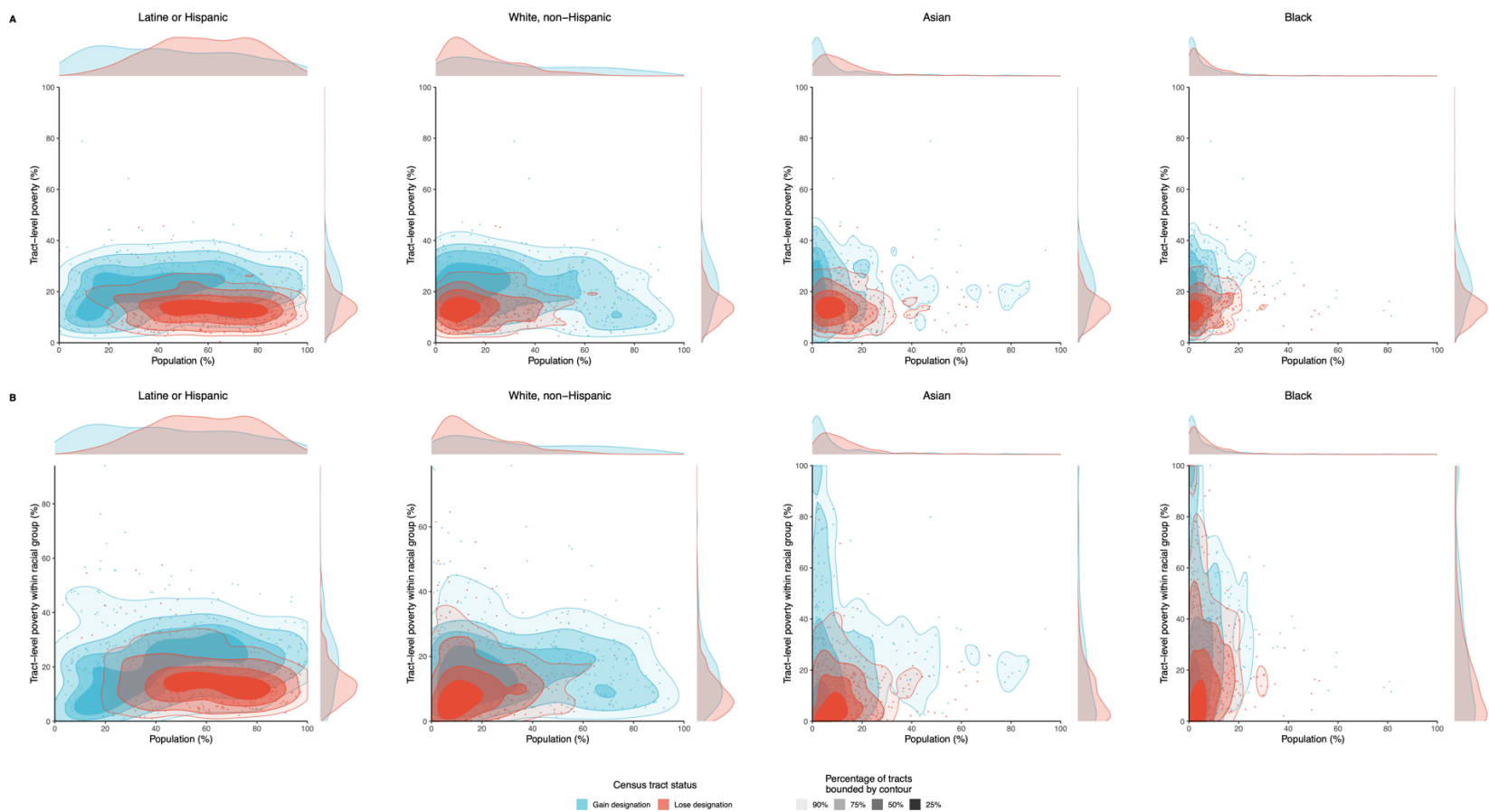

**Fig. S2. Allocative tradeoffs between poverty and populations of color.** Comparison of how algorithmically designated tracts are distributed by race and poverty across the current CalEnviroScreen model and an alternative model, among tracts that would change designation status under the alternative model. The alternative model uses a different pre-processing technique, different aggregation technique, and incorporates additional population health variables. Red densities indicate tracts that receive designation under the current model but are not designated under the alternative model. Blue densities indicate tracts gaining designation under the alternative model. Contours are calculated as the smallest regions that bound a given proportion of the data (highest density region). Dots indicate individual tracts. Similar to Fig. 2, but includes data of individual tracts (dots) and varies the y-axis on (A) poverty within racial group and (B) poverty overall.

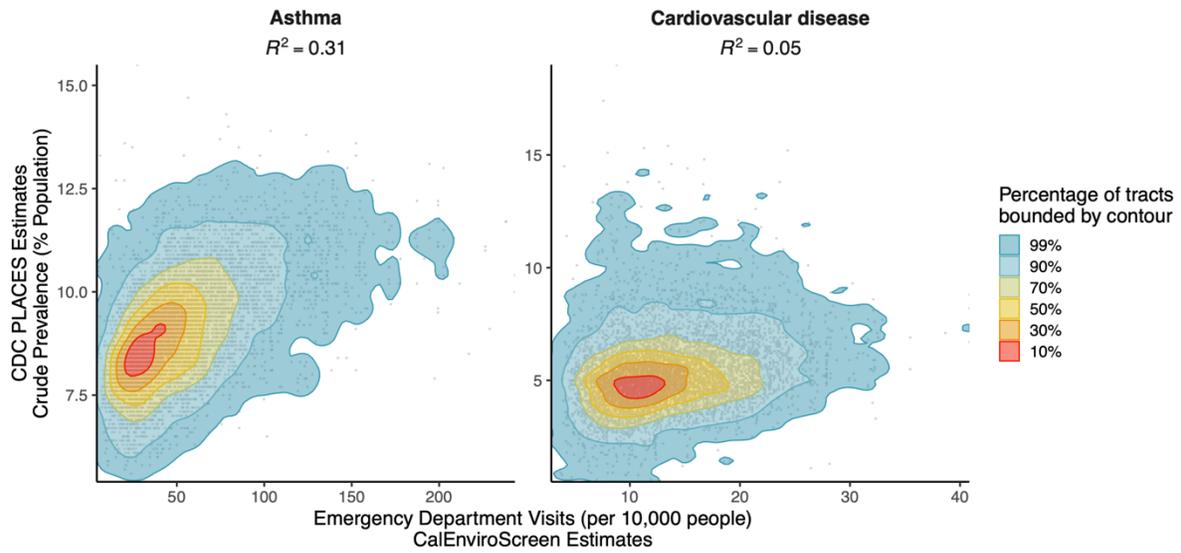

**Fig. S3. Low correlations between health indicators used by CalEnviroScreen and an alternative data source.** CalEnviroScreenCensus uses tract-level emergency room visits for asthma and heart attacks as health metrics; comparison data are indicators provided by CDC PLACES: tract-level survey data on history of asthma and coronary heart disease. Contours are calculated as the smallest regions that bound a given proportion of the data (highest density region).

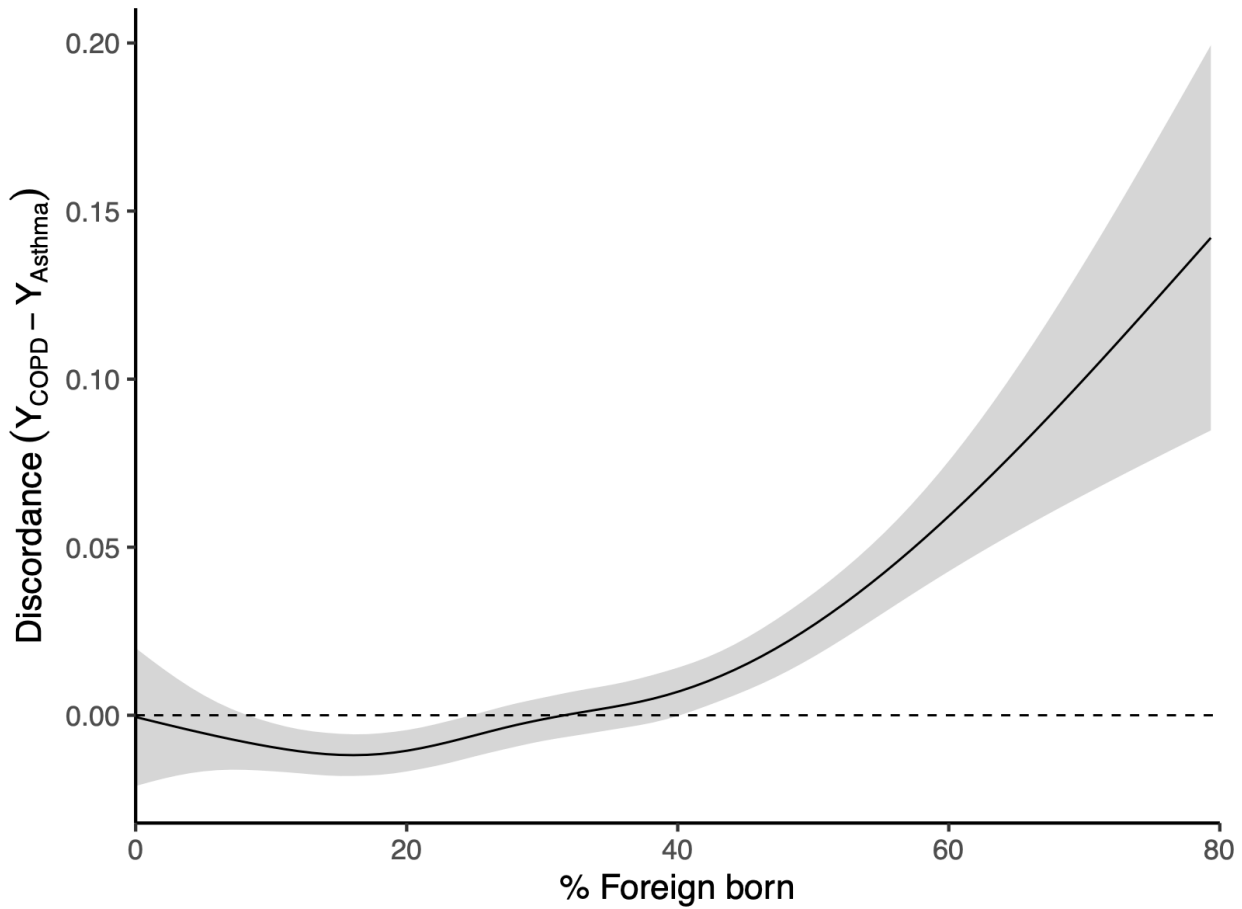

**Fig. S4. Pairwise discordance between CalEnviroScreen and an alternative model, across tracts with varying levels of foreign-born populations.** The alternative model ($Y_{COPD}$) uses survey data of chronic obstructive pulmonary disease as a measure of respiratory health compared to the current CalEnviroScreen model ($Y_{Asthma}$), which uses emergency room visits for asthma. Higher levels indicate $Y_{COPD}$ designating more tracts as disadvantaged for a given foreign born population percentage. Shaded bars indicate 95% confidence intervals, and black line indicates a smoothing spline from pointwise mean estimates of pairwise discordance. The models are comparable for tracts with fewer than a 30% foreign-born population, suggesting model bias against tracts with high immigrant populations.

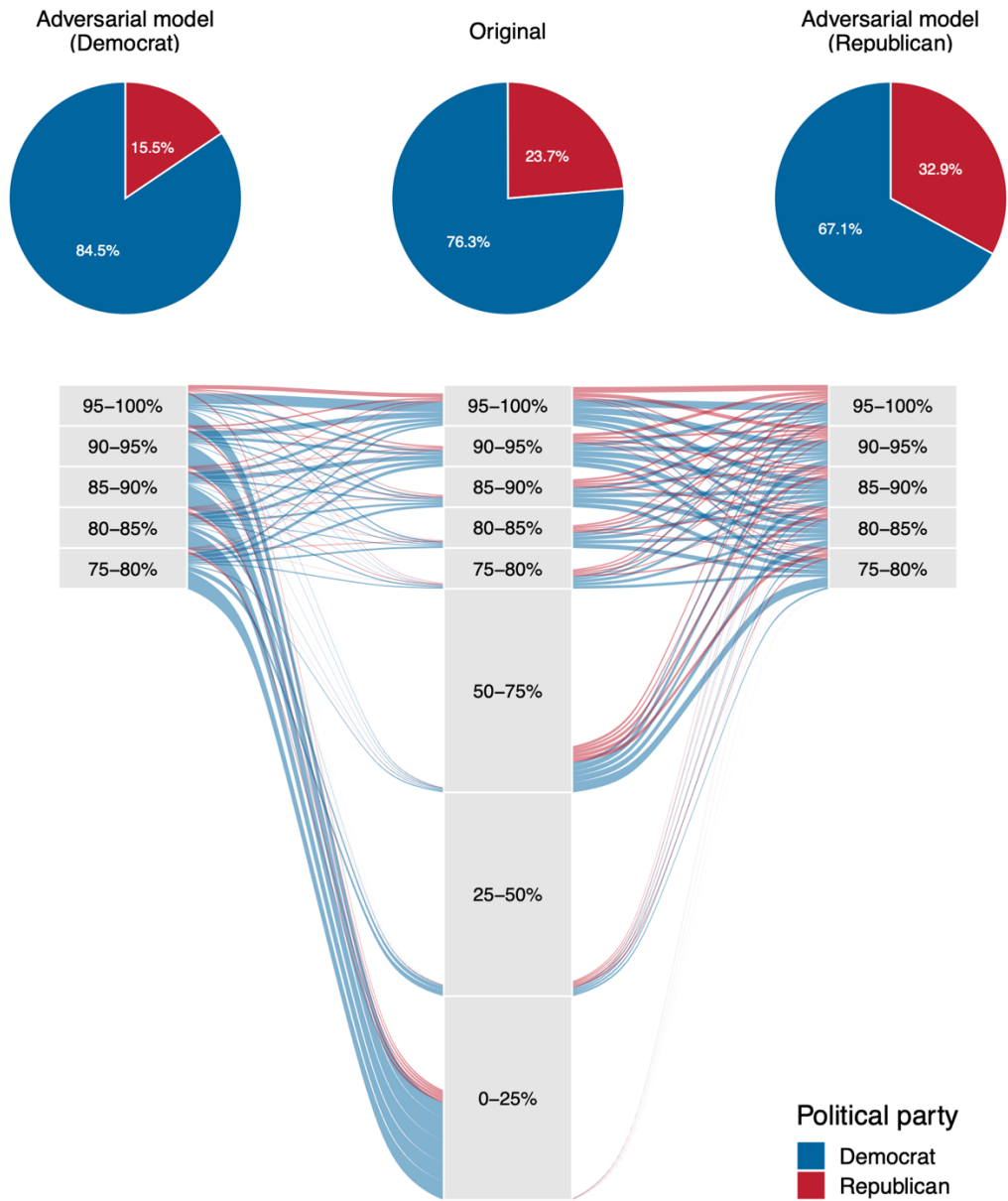

**Fig. S5. Adversarially optimized distribution of algorithmically designated tracts by political party.** Columns and flow denote distribution of tracts designated as disadvantaged by political affiliation, determined by affiliations of district assembly members. The leftmost and rightmost columns are adversarially optimized to increase Democrat and Republican tracts, respectively. Center column represents the original model. Flows between columns represent changes in binned percentile ranking of census tracts between the original and adversarial models.

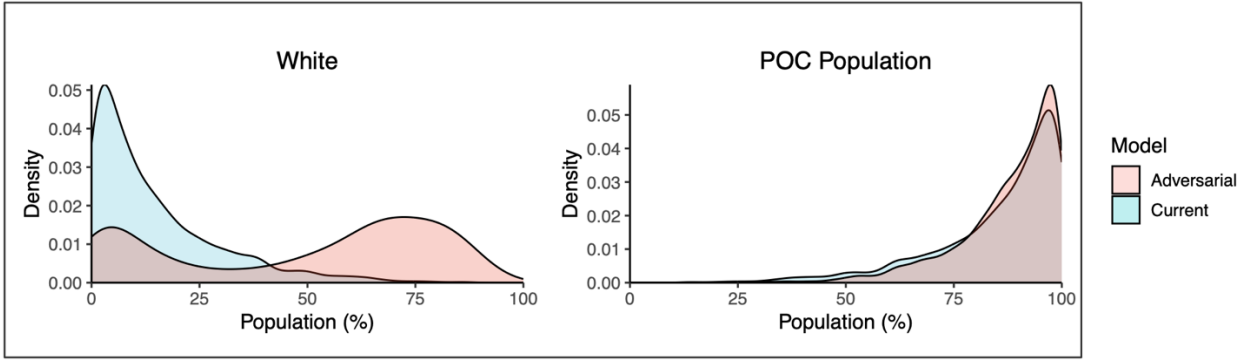

**Fig. S6. Adversarially optimized distribution of algorithmically designated tracts by race.** Blue densities represent designated tracts by the existing CalEnviroScreen model and red densities represent designated tracts by adversarially optimized models, among all algorithmically designated tracts for each model. Plot on the left compares with a model adversarially optimized to increase white populations designated for funding; plot on the right compares with a model adversarially optimized to increase populations of people of color designated for funding.

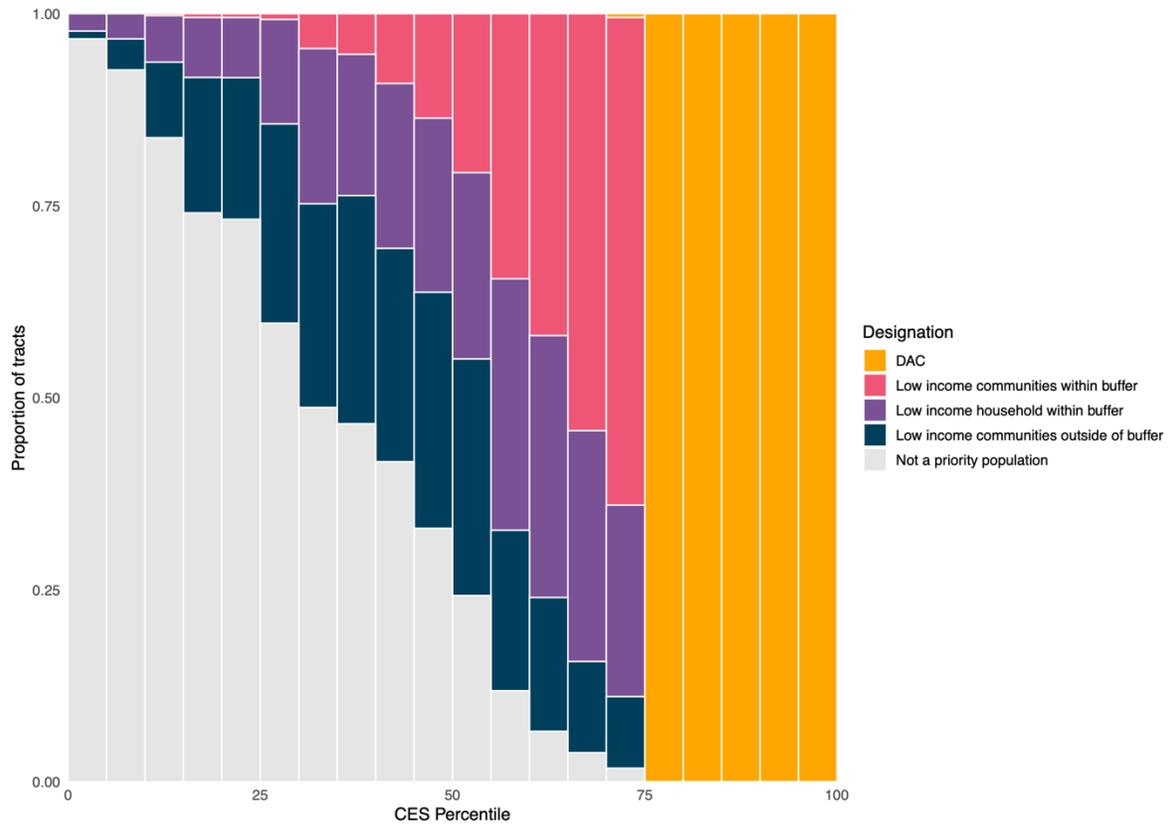

**Fig. S7. Proportion of tracts considered by California Climate Investments to be priority populations, by CalEnviroScreen Percentile.** DAC refers to disadvantaged communities as determined by scoring in the top 25% of CalEnviroScreen scores. Buffer refers to low-income communities or households that are not designated as DAC by CalEnviroScreen, but are within half a mile of a DAC tract. 25% of proceeds are earmarked for DAC tracts, 5% is earmarked for low-income households and communities statewide, and 5% is earmarked for low-income households and communities specifically within buffer regions.

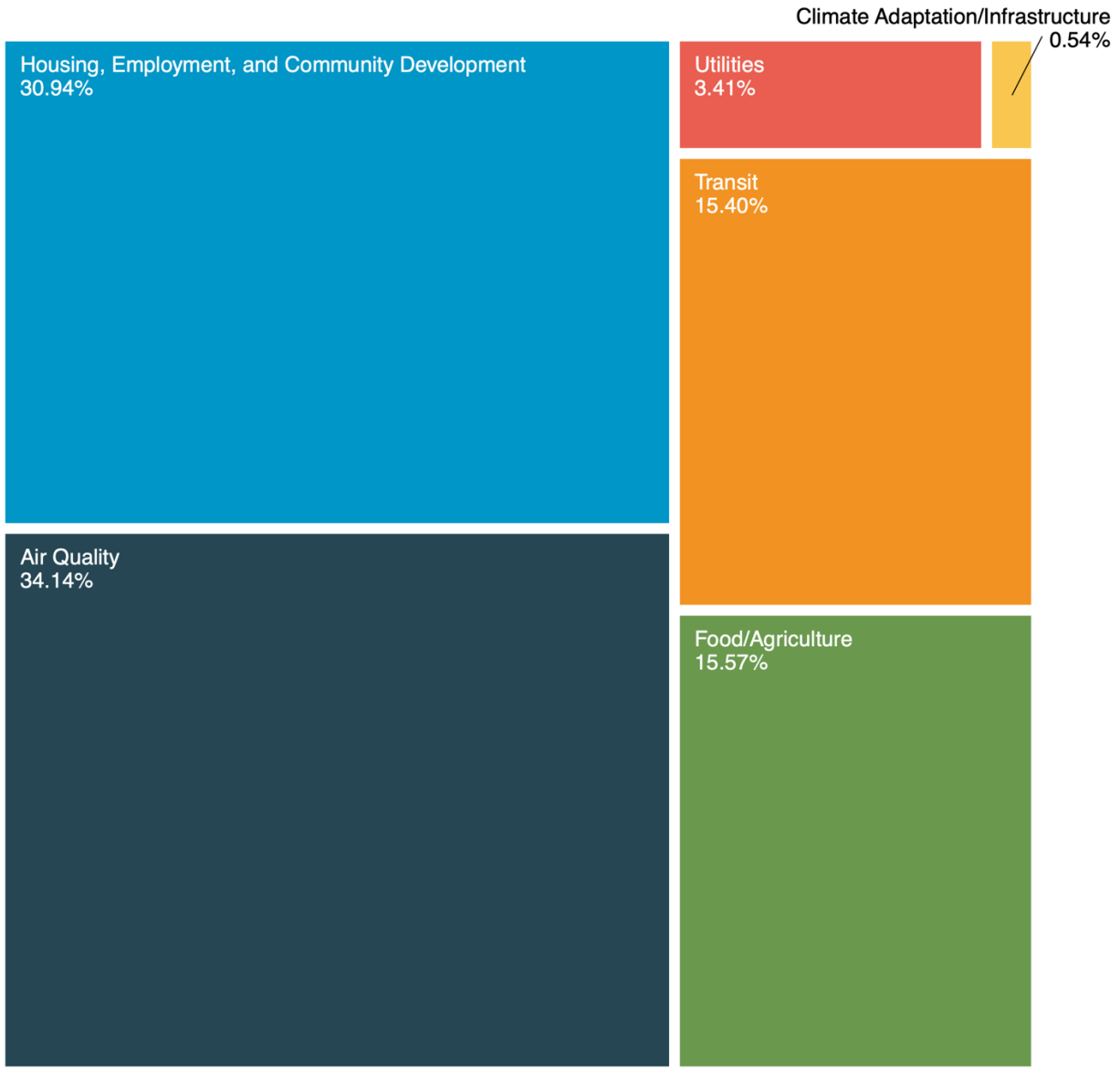

**Fig. S8. Funding allocation to tracts designated as disadvantaged communities by the CalEnviroScreen model.** Colors indicate different categories of funded projects. Cumulative funding amount is $2.6bn from 2017 to 2021.

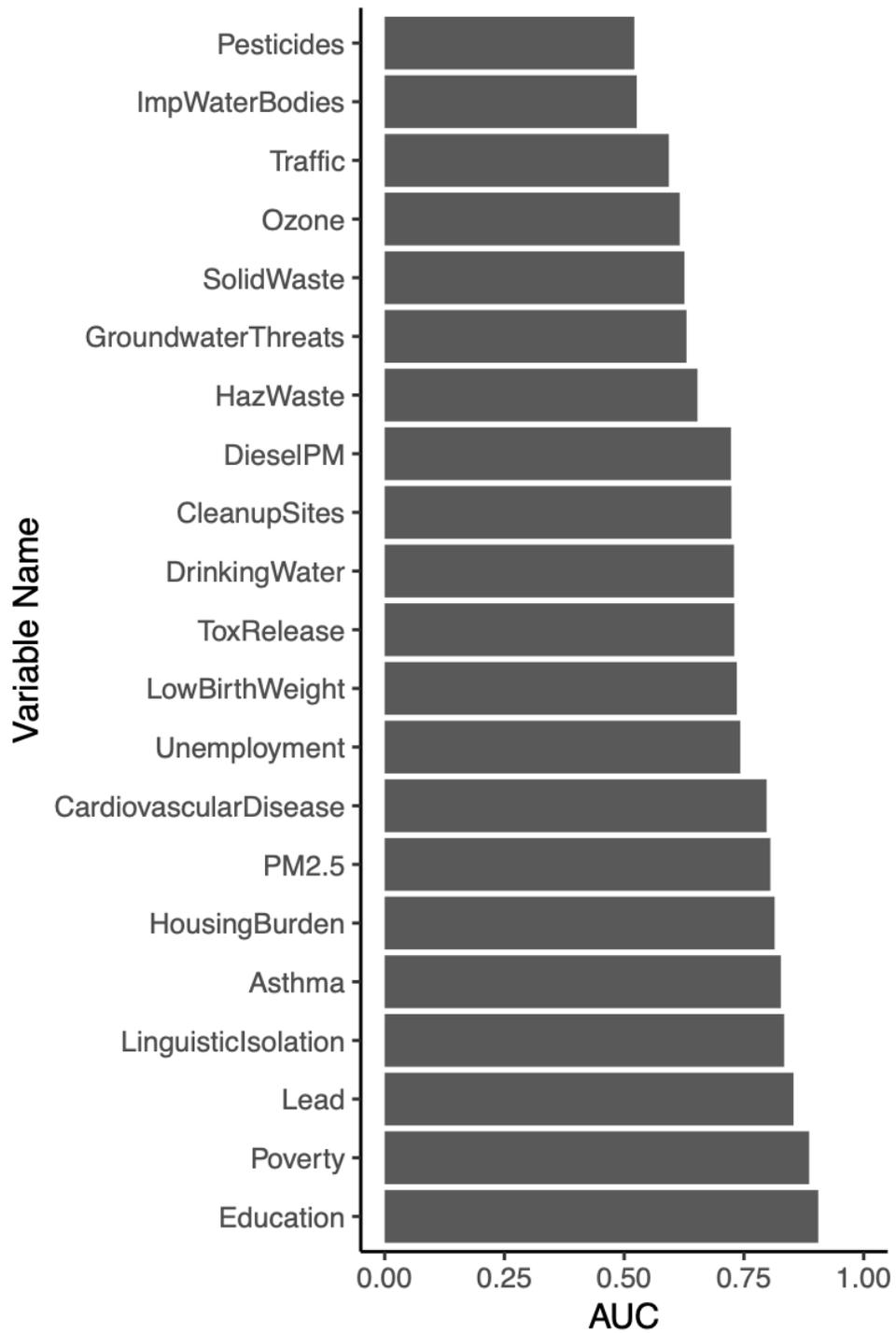

**Fig. S9. Variability in predictor strength across variables in CalEnviroScreen.** Bars denote ability to predict whether a tract will be designated as disadvantaged, measured in area under the receiver operating characteristic curve.

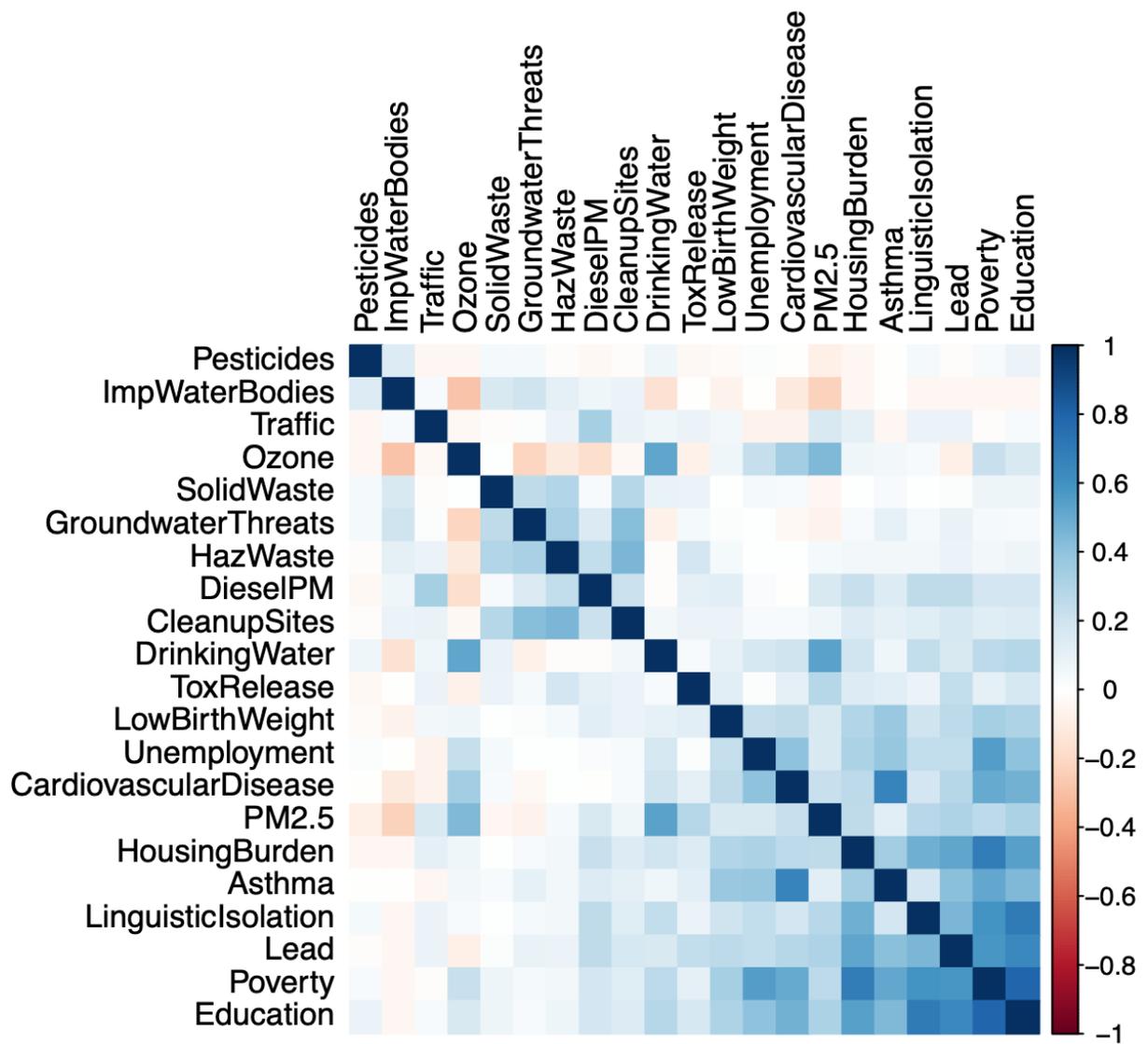

Fig. S10. Correlation plot of individual variables used by CalEnviroScreen.

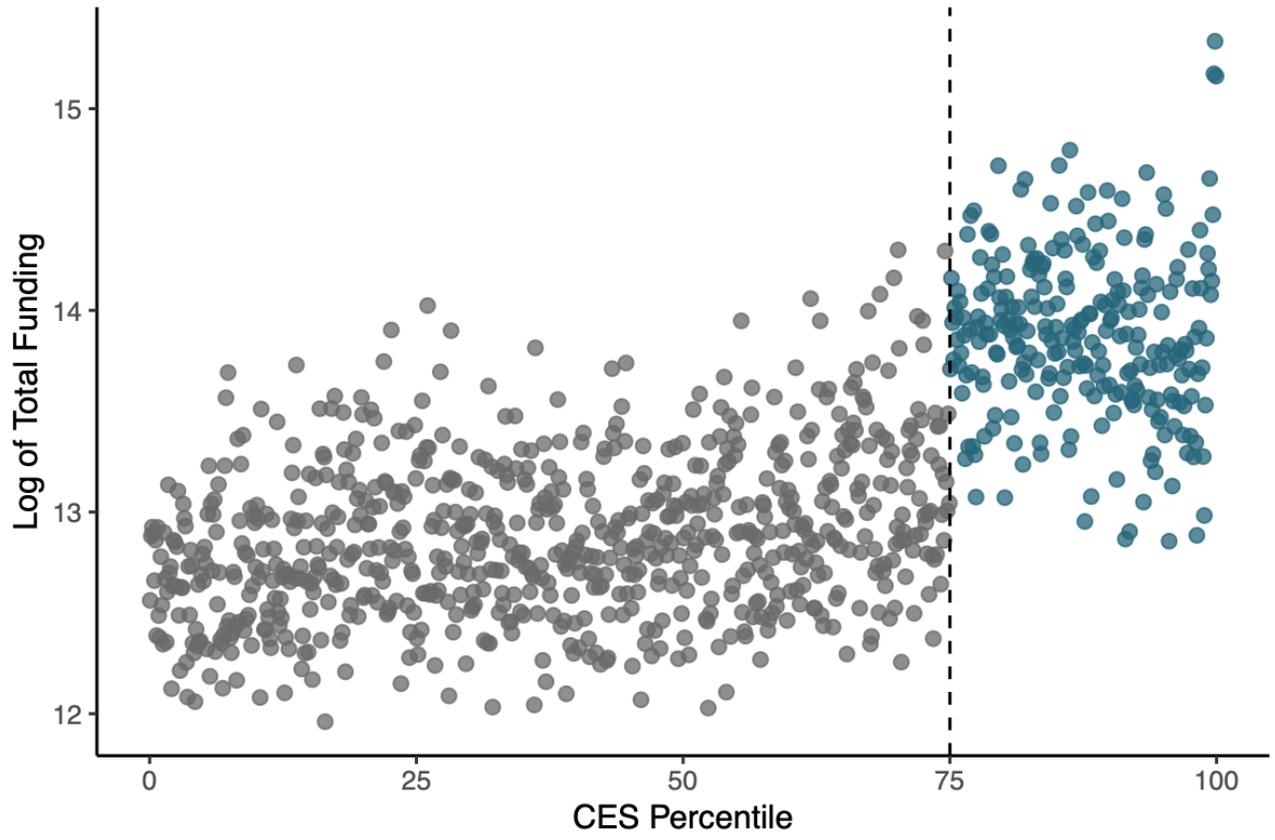

**Fig. S11. Log of total funding by CalEnviroScreen percentile.** Dots represent locally averaged funding amounts with a binwidth of 0.1. Dashed line represents the decision threshold at the top 25% of tracts.